\documentclass[twocolumn,showpacs,preprintnumbers,amsmath,amssymb]{revtex4}

\usepackage{epsfig}
\usepackage{epsf}
\usepackage{graphics}
\usepackage{graphicx}
\usepackage{amssymb}

\begin{document}

\setlength{\textheight}{23cm}
\setlength{\textwidth}{17cm}
\setlength{\oddsidemargin}{-.5cm}
\setlength{\topmargin}{-1cm}
\newcommand{\be}{\begin{equation}}
\newcommand{\ee}{\end{equation}}
\newcommand{\bq}{\begin{eqnarray}}
\newcommand{\eq}{\end{eqnarray}}
\newcommand{\kbruto}{\hbox{$k \!\!\!{\slash}$}}
\newcommand{\pbruto}{\hbox{$p \!\!\!{\slash}$}}
\newcommand{\qbruto}{\hbox{$q \!\!\!{\slash}$}}
\newcommand{\Abruto}{\hbox{$A \!\!\!{\slash}$}}
\newcommand{\dbruto}{\hbox{$\partial \!\!\!{\slash}$}}
\newcommand{\bc}{\begin{center}}
\newcommand{\ec}{\end{center}}

\title {Supergravity corrections 
to the $(g-2)_l$ factor by Implicit Regularization} 
\date{\today} 
\author{J. E. Ottoni$^{a}$ } \email []{jeottoni@fisica.ufmg.br}
\author{A. P. Ba\^eta Scarpelli$^{a,b}$ } \email []{scarp@fisica.ufmg.br} 
\author{Marcos Sampaio$^{a,c}$ } \email []{sampaio@fisica.ufmg.br, sampaio@th.u-psud.fr } 
\author{M. C. Nemes$^{a}$ } \email []{carolina@fisica.ufmg.br} 

\affiliation{$^{a}$Universidade Federal de Minas Gerais \\
Departamento de F\'{\i}sica - ICEx \\
P.O. BOX 702, 30.161-970, Belo Horizonte - MG - Brazil}

\affiliation{$^{b}$ Centro Federal de Educa\c{c}\~ao Tecnol\'ogica - MG \\
 Av. Amazonas, 7675 - Belo Horizonte - MG - Brazil}
 
\affiliation{$^{c}$ Universit\'e Paris XI -
Laboratoire de Physique Theorique\\
Orsay, 91405, France.}

\begin{abstract}
\noindent
We apply Implicit regularization in the calculation of the 
one-loop graviton and gravitino corrections to the anomalous magnetic moment
of the lepton in unbroken 
supergravity. This is an important test for any regularization method. 
We find a null result as it is expected from 
supersymmetry. We compare our results with the ones obtained by 
using Differential Regularization and Dimensional Reduction, which 
are known to preserve supersymmetry at one-loop order. 
\end{abstract}

%%%%%%%%%%%%%%%%%%%%%%%%%%%%%%%%%%%%%%%%%%%%%%%%%%%%%%%%%%%%% 
%\keywords{                 }
\maketitle 
%%%%%%%%%%%%%%%%%%%%%%%%%%%%%%%%%%%%%%%%%%%%%%%%%%%%%%%%%%%%% 

\section{Introduction}

\indent
One fundamental test for any regularization method is its applicability 
in supersymmetric theories. This is because the dynamics of the best candidates 
of fundamental theories 
is believed to respect supersymmetry. 
Searching for ideal physical calculations to perform consistency 
tests in this area, one finds among others,
the anomalous magnetic moment of the lepton in supergravity (SUGRA), 
the local version of supersymmetry (SUSY). In spite of being a 
non-renormalizable theory, gravity yields a finite result in 
one-loop correction to this observable \cite{1}. More restrictively, 
its combination with SUSY imposes that the $(g-2)_l$ factor vanishes \cite{2}, 
because there is no Pauli term in the Lagrangian of a chiral supermultiplet 
(in this context, there is a generalization to a set of sum rules 
for any charged SUSY multiplet\cite{3a}, \cite{4a}). Hence, if we want a 
regularization technique to respect supersymmetry, it must implement,  
order by order, the cancellation of the quantum corrections. 

Some methods were tested in this context \cite{3DR}, \cite{3}, \cite{3i}, \cite{3ii}, 
\cite{3d}, \cite{4d}, \cite{7d}, 
and it is well known that Dimensional 
Regularization \cite{dimr} does not provide a null result, though being finite \cite{3}. 
Dimensional Reduction \cite{dimred}, which is supersymmetric invariant, at least at one-loop 
order (some improvements are being established at higher order calculations
\cite{4}), as well as Differential Regularization \cite{1d}-\cite{7d} in its constrained version, 
were shown to give a cancellation between the contributions from the graviton and the 
gravitino sectors \cite{3DR}, \cite{3}, \cite{3d}, \cite{4d}, \cite{7d}. 

In this work, we intend to perform this important test in the context of Implicit Regularization (IR).
The basic idea behind the method is to implicitly assume
some (unspecified) regulating function as part of the integrand of
divergent amplitudes and to separate their regularization dependent
parts from the finite one. Symmetries of the model, renormalization or
phenomenological requirements  determine arbitrary parameters
introduced by this procedure \cite{6a}-\cite{13}. In fact, there is a special choice of the parameters that automatically preserves the symmetries in all non-anomalous
cases we have studied. The possibility of these parameters being fixed at the 
beginning of the calculation is desirable, since it simplifies a lot the 
application of the method.

The technique has been shown to be tailored
to treat theories with parity violating objects in integer dimensions. This is the case of
chiral and topological field theories. The ABJ anomaly  \cite{ABJ},  
and the radiative generation of a
Chern-Simons-like term,
which violates Lorentz and CPT symmetries  \cite{7}, \cite{8} are
examples
where the technique was
successfully applied. Moreover the method was shown to respect
gauge invariance
in both Abelian and non-Abelian theories at one-loop order \cite{7}, \cite{8}, \cite{9},
\cite{12}. The calculation of the $\beta$-function of the
massless Wess-Zumino model (at  three loops) was also performed as a
test  of the procedure \cite{11}. 

This paper is divided as follows: in the section II we make a short review of 
the Implicit Regularization method; in section III we establish the problem to be treated; 
in section IV, we make the calculation of the anomalous magnetic moment 
of the lepton in unbroken supergravity and compare ours with previous results. Finally, in 
section V some concluding comments are presented. 

\section{A short review of the Implicit Regularization Technique}

\indent
In order to implement the procedures prescribed by the 
Implicit Regularization technique, it is necessary that the amplitude 
be regularized. We begin by recalling the basic steps of the method:

\begin{itemize}
\item some (as yet unspecified) Regularization  is applied to the
amplitude, so as to allow for operations with the integrand.
It will be indicated by the index $\Lambda$ in the integrals; 
\item we judiciously use the identity,
\bq
\frac {1}{(p-k)^2-m^2}&=&\frac{1}{(k^2-m^2)} \nonumber \\
&-&\frac{p^2-2p \cdot 
k}{(k^2-m^2)
\left[(p-k)^2-m^2\right]},
\label{ident}
\eq
until the regularization dependent integrals (those that would be 
divergent) 
do not depend on the external momentum; 
\item we can solve the finite part which is regularization independent
and use a subtraction scheme such that the remaining regularization 
dependent integrals are eliminated. 
\end{itemize}

In order to give an example of the use of these steps, we apply the method to 
the simple logarithmically divergent one-loop amplitude below:
\be
I=\int^\Lambda \frac {d^4k}{(2\pi)^4} \frac{1}{(k^2-m^2)[(k+p)^2-m^2]}.
\ee
By applying identity (\ref{ident}) in the regularized amplitude above, 
we get
\bq
I= I_{log}(m^2)- \int_k \frac{p^2+2p\cdot k}{(k^2-m^2)^2[(k+ p)^2-m^2]},
\label{expand}
\eq
where $\int_k$ stands for $\int d^4k/(2\pi)^4$ and we have defined the basic 
one-loop logarithmically divergent object,
\be
I_{log}(m^2) = \int^\Lambda_k \frac{1}{(k^2-m^2)^2}.
\label{ilog}
\ee
Note that the second integral in eq. (\ref{expand}) is finite and, because of this, 
we do not use the superscript $\Lambda$. 
It is convenient to express the regularization dependent part, given by (\ref{ilog}), 
in terms of an arbitrary mass parameter. This becomes essential if we are treating 
massless theories (see ref. \cite{13}). It can be done by using the regularization 
independent relation 
\be
I_{log}(m^2)=I_{log}(\lambda^2)+b\ln{\left( \frac{m^2}{\lambda^2}\right)},
\label{scale}
\ee
with $b=i/(4\pi)^2$. The mass parameter $\lambda^2$ is suitable to 
be used as the renormalization group scale, as it can be seen in 
refs. \cite{11}, \cite{13}. After solving the finite part, we are left with 
\be 
I=I_{log}(\lambda^2)-bZ_0(p^2,m^2,\lambda^2),
\ee
where
\be
Z_0(p^2, m^2,\lambda^2)=\int_0^1dx\,\,\ln{\left( \frac{p^2x(1-x)-m^2}{(-\lambda^2)}\right)}.
\ee

Next we would like to address one of the essential aspects of IR: its 
application to gauge theories and the kind of constraint that should be 
introduced in order to preserve this symmetry. There is also the question 
of how to deal with anomalous theories. These aspects are suitably 
discussed in the ref. \cite{8}, but we will shortly discuss it here, 
since it is closely related to the preservation of SUSY. 
Let us consider the vacuum polarization tensor 
of QED at one-loop order. In \cite{8}, where an arbitrary routing 
in the loop momenta was used, we obtained
\bq
\Pi_{\mu \nu}&=&\Pi(p^2)(p_\mu p_\nu-p^2\eta_{\mu \nu}) \nonumber \\
&+& 4\left( \Upsilon^2_{\mu\nu}-\frac{1}{2}(k_1^2+k_2^2)\Upsilon^0_{\mu
\nu} \right. \nonumber \\
&+& \left. \frac{1}{3}(k_{1}^{\alpha}k_{1}^{\beta}+k_{2}^{\alpha}k_{2}^{\beta}
+k_{1}^{\alpha}k_{2}^{\beta})
\Upsilon^0_{\mu \nu \alpha 
\beta} \right.\nonumber \\ &-& \left.
(k_1+k_2)^{\alpha}(k_1+k_2)_{\mu}\Upsilon^0_{\nu
\alpha} \right. \nonumber \\
&-& \left. \frac{1}{2}(k_1^{\alpha}k_1^{\beta}+k_2^{\alpha}k_2^{\beta})\eta_{\mu \nu}
\Upsilon^0_{\alpha \beta}  \right), 
\label{QED}
\eq
with $p=k_1-k_2$, the external momentum. In the equation above, $\Pi(p^2)$ 
includes the divergent (regularization dependent) parts. We also have the following relations:
\be
\Upsilon_{\mu \nu}^2 \equiv \int^{\Lambda}_k
\frac{\eta_{\mu\nu}}{k^2-m^2}-
2\int^{\Lambda}_k 
\frac{k_{\mu}k_{\nu}}{(k^2-m^2)^2},
\ee
\be
\Upsilon_{\mu \nu}^0 \equiv \int^{\Lambda}_k
\frac{\eta_{\mu\nu}}{(k^2-m^2)^2}-
4\int^{\Lambda}_k 
\frac{k_{\mu}k_{\nu}}{(k^2-m^2)^3}
\label{CR1}
\ee
and
\be
\Upsilon_{\mu \nu \alpha \beta}^0 \equiv
\eta_{\{\mu \nu}\eta_{\alpha \beta \}}
\int^{\Lambda}_k
\frac{1}{(k^2-m^2)^2}
-24\int^{\Lambda}_k 
\frac{k_{\mu}k_{\nu}k_{\alpha}k_{\beta}}{(k^2-m^2)^4}.
\label{CR2}
\ee
The above differences between integrals of the same degree of divergence are 
arbitrary constants. In ref.~\cite{7}, these constants were shown to have to vanish if we want an 
amplitude independent of the choice of the momentum routing in the loops. 
This is also clear from eq.~(\ref{QED}). Moreover, by setting them to zero, 
gauge invariance is assured. Neverthenless, even if momentum routing invariance is 
violated, it is possible to find another choice that preserves gauge invariance.
Let us choose $k_2=0$ and parametrize these differences such that
\be
\Upsilon_{\mu \nu}^2=\mu ^2 \alpha_1 \eta_{\mu \nu}, \,\, 
\Upsilon_{\mu \nu}^0= \alpha_2 \eta_{\mu \nu}, \,\,
\Upsilon_{\mu \nu \alpha \beta}^0= \alpha_3 \eta_{\{\mu \nu}\eta_{\alpha \beta\}}, 
\label{CR}
\ee
so that
\be
p^\mu \Pi_{\mu \nu} = 4\left\{\mu^2\alpha_1+(\alpha_3-2\alpha_2)p^2\right\}p_{\nu}. 
\ee
It is clear that if we take $(\alpha_1,\alpha_2,\alpha_3)=(0,\alpha_2,2\alpha_2)$, 
gauge invariance is restored. The point here is the following: as long as we do not 
have any anomaly, we can simply set all those $\Upsilon$'s to zero and be sure of the 
conservation of gauge symmetry. We can call it the constrained version of Implicit 
Regularization.

Next we justify why this choice seems to be the natural one. 
As long as the symmetry is manifest in the Lagrangian of the theory, a regularization 
technique that extends the properties of regular integrals to the regularized ones 
would naturally respect this symmetry. This is the spirit in which were based the
Dimensional Regularization (see the section 3 of \cite{dimr}) and the 
Constrained Differential Regularization, for example. 
One of these important properties is the possibility of making shifts. This means 
that in regular integrals surface terms are null.
By analyzing 
the equations above, we see that if we want to eliminate all the surface 
terms, then we must choose $\alpha's=0$, since the Consistency Relations listed above 
are, in fact, proportional to surface terms. We can write:
\be
\Upsilon_{\mu \nu}^0= \int_k ^\Lambda \frac{\partial}{\partial k^\mu}
\left( \frac{k_ \nu}{(k^2-m^2)^2} \right)
\ee
and
\be
\Upsilon_{\mu \nu}^2= \int_k ^\Lambda \frac{\partial}{\partial k^\mu}
\left( \frac{k_ \nu}{(k^2-m^2)} \right).
\ee
For $\Upsilon_{\mu \nu \alpha \beta}^0$, we have
\bq
&&\int_k^\Lambda \frac{\partial}{\partial k^\beta}\left[ \frac{4k_\mu k_\nu k_\alpha}{(k^2-m^2)^3} 
- \frac{k_\alpha \eta_{\mu \nu} +k_\mu \eta_{\nu \alpha} +k_\nu \eta_{\mu \alpha}}
{(k^2-m^2)^2}\right] \nonumber \\
&&=\eta_{\{\mu \nu}\eta_{\alpha \beta\}}(\alpha_3-2 \alpha_2),
\label{sup2}
\eq
where we have made use of equations (\ref{CR1}), (\ref{CR2}) and (\ref{CR}).
If the expression of eq.(\ref{sup2}) vanish, we find the alternative 
condition to respect gauge invariance in the QED vacuum polarization tensor discussed
above. In this particular case, it was not necessary to make also $\alpha_2=0$, but 
this situation will not hold in general. 
An important remark is that the parameters that come from logarithmically divergent 
integrals are finite. As surface terms, they can be calculated 
by symmetric integration. The procedure of making them zero corresponds, indeed, 
to automatically add local symmetry restouring counterterms in order to cancel surface terms. 
In Dimensional Regularization, they are null, as it would be all surface terms. 
This principle can be applied to higher order calculations \cite{eds}. 
At n-loop order, besides the  
CR of all the previous orders, some others will be needed. Generally, a Consistency 
Relation with m Lorentz indices can be obtained from the integral of 
the m-th order derivative of the corrected  fermion   
internal line. 

The situation changes completely if we are treating anomalous processes 
as, e.g., the anomalous pion decay (axial--vector--vector triangle diagram) \cite{ABJ}. 
In this case, in order to have a ``democratic'' choice of the preserved identities,
one is forced to violate momentum routing invariance, as shown in ref.~\cite{8}.

\section{SUGRA Lagrangian and the statement of the problem}

We are interested in the coupling of supergravity with matter, which has been vastly described 
in literature \cite{matgra}. We are considering the linearized interaction 
Lagrangian density of superQED-SUGRA, which in Minkowski space is given by 
\bq
{\cal L}_I&=& -e\bar \Psi \Abruto \Psi -\left[ieA^\mu \varphi^\dagger_L\overleftrightarrow
\partial_\mu \varphi_L \right. \nonumber \\
&-& \left. e \sqrt 2(\bar \lambda \varphi ^\dagger_LP_L\Psi +h.c.) +(L \leftrightarrow R)
\right] \nonumber \\
&-& \frac{\kappa}{4} h^{\alpha \beta} \left[(i \bar \Psi(\gamma_\alpha\partial_\beta 
+\gamma_\beta \partial_\alpha)\Psi + h. c.) \right. \nonumber \\
&-&\left. 2e \bar \Psi(\gamma_\alpha A_\beta + \gamma_\beta A_\alpha)\Psi \right] \nonumber \\
&-&\frac{\kappa}{\sqrt 2} \left[ \bar \chi^\nu P_L(i \dbruto-m)\varphi^\dagger_L 
\gamma_\nu \Psi \right. \nonumber \\
&+& \left. e \bar \chi^\nu P_L\Abruto \varphi^\dagger_L \gamma_\nu \Psi + h.c. \right] 
+ (L \leftrightarrow R) \nonumber \\
&+& \kappa \left[ h^{\alpha \beta}(F_{\alpha \mu}F_\beta^{\,\mu}-
\frac 14 \eta_{\alpha \beta}F_{\mu \nu}F^{\mu \nu}) \right. \nonumber \\
&+& \left. (\frac i8 \bar \lambda \gamma^\nu[\dbruto,\Abruto]\chi_\nu + h.c.) \right],
\eq
where $\overleftrightarrow \partial = \partial - \overleftarrow \partial$, 
$P_{R,L}=\frac 12 (1 \pm \gamma_5)$ are the usual chiral projectors and 
we have $\kappa^2=8 \pi G_N$, $G_N$ being the Newton's gravitational constant.
The following notation is used for the fields: 
$\Psi$ for the lepton, $A_\mu$ for the photon, $h_{\mu \nu}$ for the 
graviton, $\varphi_{L,R}$ for the slepton, $\lambda$ for the photino and $\chi_\mu$ 
represents the gravitino.

In our problem, the determination of the $(g-2)_l$ factor to one-loop order, we are 
interested in the quantum corrections to the vertex lepton-photon-lepton. By Lorentz 
invariance and current conservation, we must obtain something of the form
\be
eA_\mu \bar \Psi \left[ \gamma^\mu F_1(q^2)+i \sigma^{\mu \nu}\frac{(p-p')_\nu}{2m}F_2(q^2)
\right]\Psi,
\label{QC}
\ee 

\begin{widetext}

\begin{figure}
\unitlength1cm
\centerline{\hbox{
   \epsffile{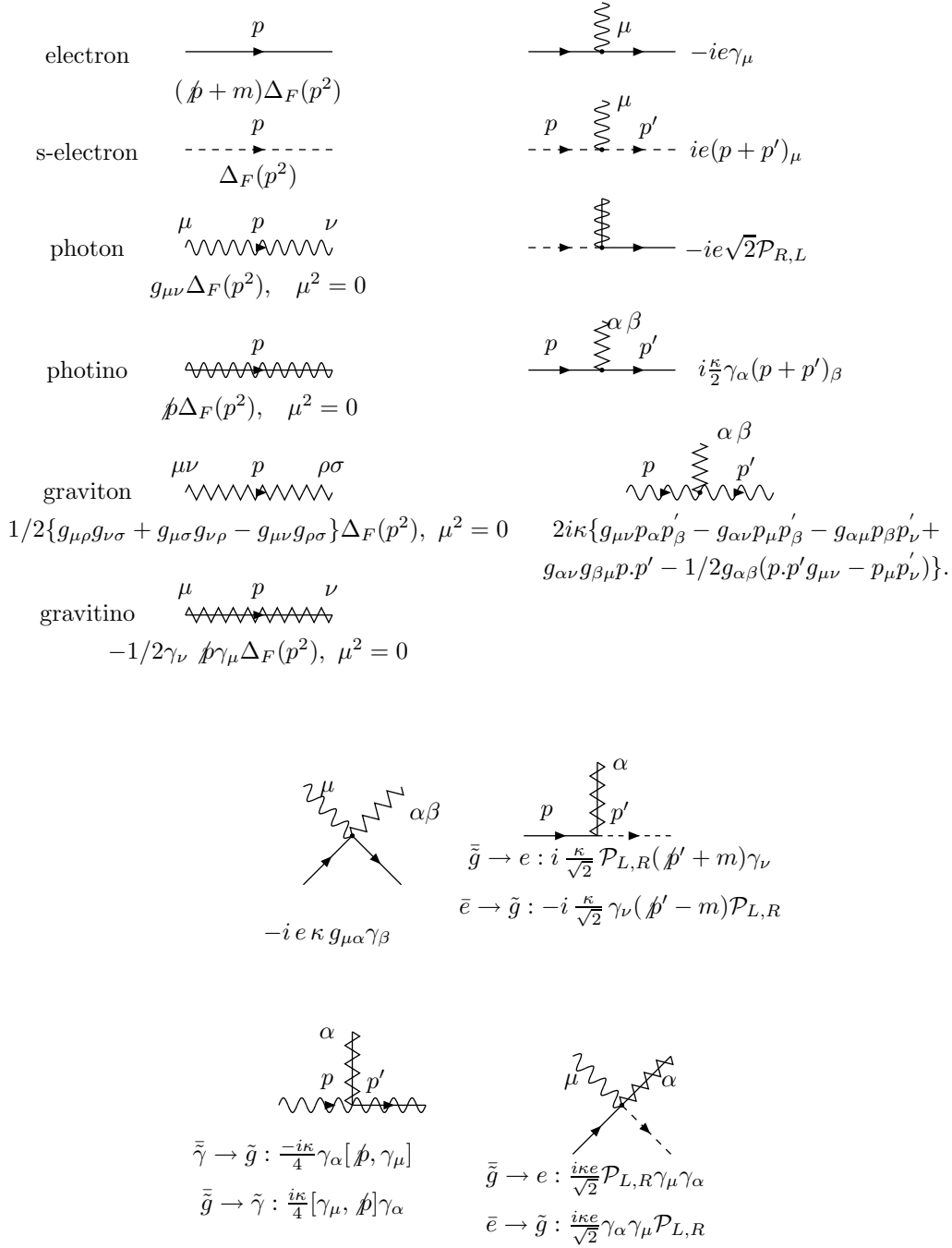}
     }
  }
\caption[]{\label{fig1} Momentum space Feynman rules} 
\end{figure}

\end{widetext}

\begin{widetext}

\begin{figure}
\unitlength1cm
\epsffile{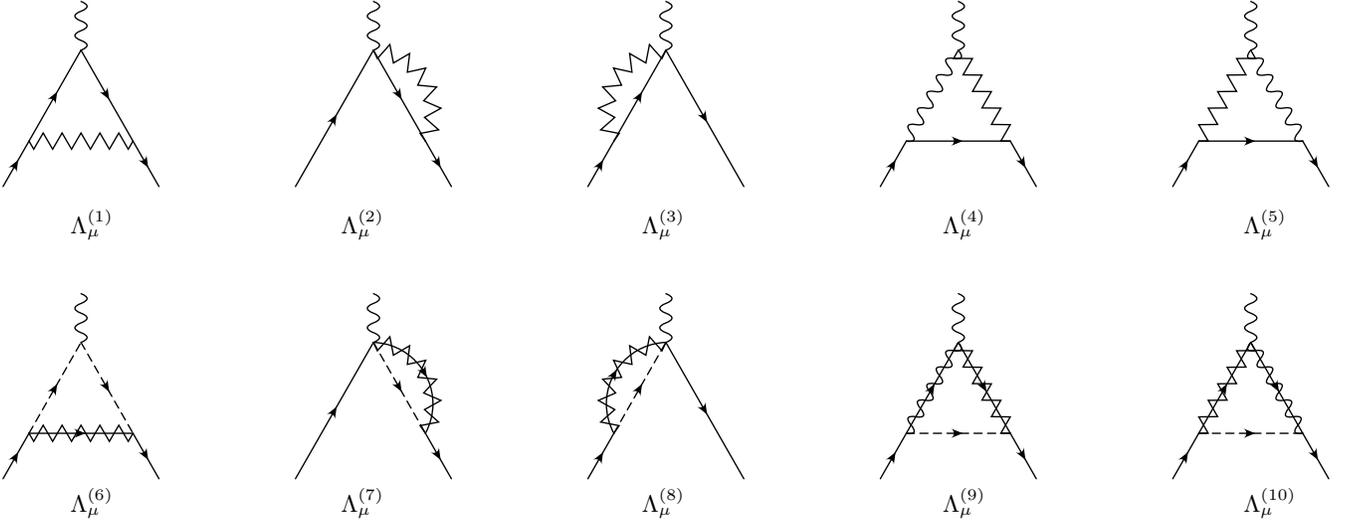}
\caption[]{\label{fig2} SUGRA-QED contributions to the $e e \gamma$ vertex}
\end{figure}

\end{widetext}

where $\sigma^{\mu \nu}= \frac i2[\gamma^\mu, \gamma^\nu]$ and $q=p-p'$ is the photon momentum.  
So, the on-shell form factor $F_2(q^2=0)$ gives us the $(g-2)_l$ factor. By using Gordon 
decomposition and the Clifford algebra, $\{\gamma^\mu, \gamma^\nu\}=2\eta^{\mu \nu}$, 
eq. (\ref{QC}) can be suitably arranged such that we have
\be
eA_\mu \bar \Psi \left[ \gamma^\mu (F_1+F_2)+i \frac{(p+p')^\mu}{2m}F_2
\right]\Psi.  
\ee
Since we are interested only in $F_2$, we will not consider the terms in $\gamma^\mu$ 
in the calculations to be done. The diagrams which contribute to the one-loop correction 
to the lepton-photon-lepton vertex are displayed in fig. 2. So, we have 
\be
\Lambda_\mu= \sum_{j=1}^{10} \Lambda_\mu^{(j)}.
\ee

\section{The $(g-2)_l$ factor by Implicit Regularization}

We will now determine the anomalous magnetic moment of the lepton as 
stated in the previous section.  
We will explicitly calculate here the contribution $\Lambda_\mu^{(2)}$, 
represented by the second diagram of figure 2. 
Using the Feynman rules of figure 1, we can write
\bq
\Lambda_\mu^{(2)}&=&
\int_k^\Lambda \frac
{1}{D}(\frac{i\kappa}{2})\gamma_{\alpha}i(\pbruto ' - \kbruto  +
m)(2p' - k)_{\beta}  \nonumber \\
&\times&\frac {i}{2}(\eta^{\alpha \rho}\eta^{\beta \sigma} +
\eta^{\beta \rho}\eta^{\alpha \sigma} - \eta^{\alpha \beta}\eta^{\rho
\sigma})  \nonumber \\
&\times&(-ie\kappa \eta_{\rho \mu} \gamma_{\sigma}) \nonumber \\
&=& \frac {-e\kappa^{2}}{4}\int_k^\Lambda \frac{N_\mu}{D}, 
\label{d2}
\eq
where
\be
D=[(p'-k)^2-m^2]k^2
\ee
and
\bq
N_\mu&=&\gamma_{\mu}(\pbruto ' - \kbruto  + m)(2\pbruto ' -
\kbruto ) \nonumber \\
&+& \gamma_{\alpha}(\pbruto ' - \kbruto  
+m)\gamma^{\alpha}(2p' - k)_{\mu} \nonumber \\
&-& (2\pbruto ' - \kbruto )(\pbruto '
- \kbruto  + m)\gamma_{\mu}.
\eq
If the on-shell condition and the Clifford algebra are used,
we get
\bq
N_\mu&=&(-4m^{2} + 4m\kbruto  - 2(p \cdot k))\gamma_{\mu} \nonumber \\
&+& (8m + 6\kbruto )p'_{\mu} + (-6m - 2\kbruto )k_{\mu},
\eq
which, when substituted in eq. (\ref{d2}), gives
\bq
\Lambda_\mu^{(2)}&=&\frac {-e\kappa^{2}}{4}[8mp'_{\mu}I(p') +
6\gamma^{\alpha}p'_{\mu}I_{\alpha}(p') \nonumber \\
&-& 6mI_{\mu}(p') -2m\gamma^{\alpha}I_{\alpha \mu}(p')+A \gamma_\mu].
\label{d22}
\eq
In the expression above, $A$ represents all the terms that multiply
$\gamma_\mu$ which are irrelevant for the problem in question. There are also divergent integrals
\be
I,I_\mu, I_{\mu \nu}(p')=\int_k^\Lambda 
\frac {1,k_\mu, k_\mu k_\nu}{k^2[(p'-k)^2-m^2]},
\ee  
which, on-shell evaluated, yield the following results:
\be
I=I_{log}(\lambda^2)+ b \ln{\left(\frac{\lambda^2}{m^2}\right)}+ 2b,
\ee
\be
I_\mu=\frac{p'_\mu}{2} \left \{  I_{log}(\lambda^2)+ 
b \ln{\left(\frac{\lambda^2}{m^2}\right)}+b + \alpha_2\right\}
\ee
and
\bq
I_{\mu \nu}&=&\frac 13 p'_\mu p'_\nu \left\{
I_{log}(\lambda^2)+ b \ln{\left(\frac{\lambda^2}{m^2}\right)}
+ \frac 23 b  +\alpha_3\right\} \nonumber \\
&+& \mbox{$\eta_{\mu \nu}$ terms},
\eq
with the $\alpha_i$'s being the arbitrary parameters defined in eq. (\ref{CR}). 
If we substitute the results of the integrals in eq. (\ref{d22}), we obtain 
\bq
\Lambda_\mu^{(2)}&=&\frac {-e\kappa^{2}}{4}\left\{ \frac{22}{3} I_{log}(\lambda^2)
+\frac{22}{3}b\ln{ \left( \frac{\lambda^2}{m^2}\right)} \right. \nonumber \\
&+& \left. \frac {140}{9}b -\frac 23 \alpha_3 \right\}m p'_\mu   
+ A \gamma_\mu.
\eq
This amplitude, when added up to $\Lambda_\mu^{(3)}$, obtained by the 
interchange $p \leftrightarrow p'$, yields for the coefficient of $(p+p')_\mu$:
\be
\frac {-e\kappa^{2}}{4}m\left\{ \frac{22}{3} I_{log}(\lambda^2)
+\frac{22}{3}b\ln{ \left( \frac{\lambda^2}{m^2}\right)}+
\frac {140}{9}b -\frac 23 \alpha_3 \right\}.
\ee
So, given that $\kappa^2=8 \pi G_N$, 
$b=i/(4\pi)^2$ and, by the definition of $F_2$, we must multiply the 
coefficient of $(p+p')_\mu$ by $2m/(ie)$, we find 
\bq
F_2^{(2+3)}&=&\frac{G_Nm^2}{\pi} \left\{-\frac{11}{6} \left(\frac 1b I_{log}(\lambda^2)+ 
\ln{ \left( \frac{\lambda^2}{m^2}\right)}\right) \right. \nonumber \\
&-& \left. \frac{35}{9} - \frac 83 i \pi^2 \alpha_3 \right\}.
\eq 
The same procedure is applied to obtain the remaining contributions to the 
total vertex $\Lambda_\mu$. The on shell results in terms of the integrals defined 
in the appendix are:
\bq
\tilde \Lambda_\mu^{(1)}&=&\frac{\kappa^2 e}{8} \left \{ 2(2p'_\mu - m \gamma_\mu) \gamma^\alpha 
(4m^2 J_\alpha+I_\alpha(p)) \right. \nonumber \\
&+& \left. 2\gamma^\alpha (2p_\mu - m \gamma_\mu)(4m^2 J_\alpha+I_\alpha(p')) 
\right. \nonumber \\
&-& \left. 24m^2\gamma^\alpha J_{\alpha \mu}
+8m(2I_\mu(p,p')-I_\mu(p)-I_\mu(p')) \right. \nonumber \\
&-& \left. 8m(p+p')_\mu I(p,p') \right\},
\eq
\bq
&&\tilde \Lambda_\mu^{(4)}+\tilde \Lambda_\mu^{(5)}= \frac{\kappa^2 e}{2}
\left\{- p_{\mu}\gamma^{\alpha}(I_{\alpha}(p) -
I_{\alpha}(p')) \right. \nonumber \\
&&  \left. +\gamma^{\alpha} p'_{\mu}(I_{\alpha}(p) -
I_{\alpha}(p')) - 2mp'_{\mu}(I(p) - I(p')) \right. \nonumber \\
&& \left. + m(\gamma^{\alpha},\gamma_{\mu})(I_{\alpha}(p) - I_{\alpha}(p')) \right. \nonumber \\
&& \left. +2(p+p')_{\mu}(\gamma^{\alpha}I_{\alpha}(p,p') - mI(p,p')) \right. \nonumber \\
&&\left. +2p_{\mu}(I(p) + I(p'))m - 4p_{\mu}mI(p,p') - 8m^{3}p_{\mu}J' \right. \nonumber \\
&& \left. -2m(I_{\mu}(p) + I_{\mu}(p')) + 4mI_{\mu}(p,p') + 8m^{3}J'_{\mu} \right. \nonumber \\
&& +\left. 8m^{2}p_{\mu}\gamma^{\alpha}J'_{\alpha} -
8m^{2}\gamma^{\alpha}J'_{\alpha \mu} + p \leftrightarrow p'\right\},
\eq
\bq
\tilde \Lambda_\mu^{(6)}&=& -\frac{\kappa^2e}{4} \left\{ 
4(p+p')_\mu[m(I(p)+I(p')) \right. \nonumber \\
&+& \left. \gamma^\alpha(I_\alpha(p)+I_\alpha(p'))] \right. \nonumber \\
&-& \left. 8[m(I_\mu(p)+I_\mu(p')) +\gamma^\alpha(I_{\mu \alpha}(p)+I_{\mu \alpha}(p'))]
\right. \nonumber \\ 
&+& \left. 12m^2 \gamma^\alpha J_\alpha (p+p')_\mu -24 m^2 \gamma^\alpha J_{\mu \alpha},
\right\}
\eq  
\bq
\tilde \Lambda_\mu^{(7)}+\tilde \Lambda_\mu^{(8)}&=& 2\kappa^2e
\left\{ p'_\mu \gamma^\alpha I_\alpha(p')-mI_\mu(p') \right. \nonumber \\
&-& \left. \gamma^\alpha I_{\mu \alpha}(p') + p \leftrightarrow p' \right\},
\eq
\bq
\tilde \Lambda_\mu^{(9)}+ \tilde \Lambda_\mu^{(10)}&=& \kappa^2 e 
\left\{ 2p'_\mu[ mI(p')-2mI(p,p') \right. \nonumber \\ 
&+& \left. \gamma^\alpha I_\alpha(p,p')+2m^2\gamma^\alpha J'_\alpha -2m^3 J] 
\right. \nonumber \\ 
&+& \left. (p-p')_\mu [\gamma^\alpha I_\alpha(p,p')-m I(p,p')] \right. \nonumber \\ 
&+& \left. m \gamma^\alpha(I_\alpha(p)-I_\alpha(p')) \gamma_\mu +2m[ -I_\mu(p) 
\right. \nonumber \\
&+& \left. I_\mu(p,p') +2m^2J'_\mu -2m\gamma^\alpha J'_{\mu \alpha}] 
\right. \nonumber \\ 
&+& \left. p \leftrightarrow p' \right\}.
\eq
In the expressions above, the tilde indicates that we have explicitly 
neglected terms with coefficients $\gamma_\mu$. Adding all the on-shell 
results obtained above we get for the contributions to the $F_2$ form factor:
\bq
F^{(1)}_2&=& \frac{G_Nm^2}{\pi} \left\{-\frac 16 \left( \frac 1b I_{log}(\lambda^2)
+\ln{ \left( \frac{\lambda^2}{m^2}\right)}\right) \right. \nonumber \\
&-& \left. \frac{29}{18} -2i \pi^2\left(\frac 23 \alpha_3 +6\alpha_2\right) \right\} \\
F^{(4+5)}_2&=& \frac{G_Nm^2}{\pi} \left\{2 \left(\frac 1b I_{log}(\lambda^2)+ 
\ln{ \left( \frac{\lambda^2}{m^2}\right)}\right) \right. \nonumber \\
&+& \left. 6 -64 i \pi^2 \alpha_2\right\} \\
F^{(6)}_2&=& \frac{G_Nm^2}{\pi} \left\{-\frac{4}{3} \left(\frac 1b I_{log}(\lambda^2)+ 
\ln{ \left( \frac{\lambda^2}{m^2}\right)}\right) \right. \nonumber \\
&-& \left. \frac{37}{18} -\frac {32}{3} i \pi^2 \alpha_3 \right\} \\
F^{(7+8)}_2&=& \frac{G_Nm^2}{\pi} \left\{-\frac{2}{3} \left(\frac 1b I_{log}(\lambda^2)+ 
\ln{ \left( \frac{\lambda^2}{m^2}\right)}\right) \right. \nonumber \\
&-& \left. \frac{4}{9} +\frac {32}{3} i \pi^2 \alpha_3 \right\}  \\
F^{(9+10)}_2&=& \frac{G_Nm^2}{\pi} \left\{2 \left(\frac 1b I_{log}(\lambda^2)+ 
\ln{ \left( \frac{\lambda^2}{m^2}\right)}\right) \right. \nonumber \\
&+& \left. 2 -64 i \pi^2 \alpha_2 \right\}.
\eq
So, we have in the graviton sector,
\bq
F^{(1)}_2+ F^{(2+3)}_2+ F^{(4+5)}_2 &=&\frac{G_Nm^2}{\pi} \left\{ \frac 12 \right. \nonumber \\
&-& \left. 4i \pi^2(19\alpha_2+ \alpha_3)\right\} ,
\eq
which is finite for any value of $\alpha_2$ and $\alpha_3$, since they are 
finite surface terms.
In the gravitino sector, we obtain 
\be  
F^{(6)}_2+ F^{(7+8)}_2+ F^{(9+10)}_2=\frac{G_Nm^2}{\pi} \left\{ -\frac 12
-64 i \pi^2 \alpha_2 \right\}.
\ee
For the total one-loop correction, we have
\be
\sum_{i=1}^{10} F^{(i)}=\frac{G_Nm^2}{\pi} \left\{-4i\pi^2(35\alpha_2+\alpha_3) \right\}.
\ee
As required by supersymmetry, the contributions from the two sectors cancel out if 
we set $\alpha_2=\alpha_3=0$. Of course this is not the only choice for this calculation. 
For instance, $\alpha_3=-35\alpha_2$ is compatible with
supersymmetry. Although for any value of the $\alpha's$ we do not have 
terms proportional to the photon momentum (this would explicitly violate 
gauge symmetry) this choice would violate gauge invariance, as we 
have seen in section II. On the other hand, the choice $\alpha_i=0$ enforces 
momentum routing invariance and consequently gauge invariance, as long as 
we do not have anomalies. 

We would like to compare our results with the ones obtained by using 
Differential Regularization and 
Dimensional Reduction (DRed). The SUGRA $(g-2)_l$ factor by Constrained Differential Regularization (CDR) was 
calculated in \cite{3d}, \cite{4d} and \cite{7d} (in ref. \cite{7d} some 
minimal corrections are 
made in the results of refs. \cite{3d} and \cite{4d}). 
We can say that we have got equivalent results. If we choose a scheme of subtraction such 
that the $I_{log}(\lambda^2)$ divergences are eliminated and rescale our arbitrary parameter 
so that $\bar \lambda^2 = \exp(2) \lambda^2$, the results become exactly the same for each independent diagram. 
It is in progress a work in which the equivalence between IR and CDR is shown \cite{irdif}. In the 
context of Implicit Regularization, the relation between bubble and triangular diagrams is obtaining 
by cancelling factors of the numerator with factors of the denominator in the integrand. The 
rules of CDR take care of momentum space surface terms. As an example, let us analyze 
the triangular basic function, $T_{mmm}[\partial_\mu \partial_\nu]$, of CDR in momentum space. We follow 
the same steps: it is decomposed into a trace and a traceless part plus an arbitrary local term:
\bq
\hat T^R[k_\mu k_\nu]&=& \hat T[k_\mu k_\nu -\frac 14 \eta_{\mu \nu}k^2]+ \frac 14 \eta_{\mu \nu} 
\hat T^R[k^2] \nonumber \\
+c \eta_{\mu \nu}.
\eq
We will fix this local term in such a way that it cancel the surface term that comes from 
the traceless part. By using equation (\ref{ident}) and the identity $k^2=(k^2-m^2)+ m^2$ in this term, we get 
\be
-\frac 14 \alpha_2 \eta_{\mu \nu} + \mbox{nonambiguous terms}.
\ee
We calculate the first term by symmetric integration ($k_\mu k_\nu \to k^2 \eta_{\mu \nu}/4$):
\bq
&&-\frac 14\left\{\int^{\Lambda}_k
\frac{\eta_{\mu\nu}}{(k^2-m^2)^2}-
4\int^{\Lambda}_k 
\frac{k_{\mu}k_{\nu}}{(k^2-m^2)^3}\right\} \nonumber \\
&& = \frac{m^2}{4}\eta_{\mu \nu}\int_k \frac{1}{(k^2-m^2)^3}= -\frac{i}{128 \pi^2}\eta_{\mu \nu}.
\eq
So, in order to cancel this surface term, $c= i/(128 \pi^2)$, just like in \cite{6d} 
(the $i$ factor is due to the fact that we work in Minkowski space). 

Concerning the calculation by Dimensional Reduction \cite{3DR}, \cite{3}, we can say that 
the results of the present paper are identical for each diagram. One only needs 
to calculate $I_{log}(\lambda^2)$ 
by dimensional integration in order to check this. 
The fact that in DRed the gamma matrices are 
treated in four dimensions implies, in the case of one-loop calculations, that the two procedures are 
equivalent. This discussion is closely related to the one carried out in ref. \cite{7d}.

We would also like to remark that there is no possibility of reproducing 
the result of Dimensional Regularization, diagram by diagram, with a global choice of the arbitrary constants. 

\section{Conclusions}

In this paper we have performed one important test with the Implicit Regularization method: 
we applied this technique to obtain a physical observable in a 
supersymmetric gauge theory. As it was expected, the cancellation between the graviton and 
the gravitino contributions to the anomalous magnetic moment of the lepton in SUGRA-QED is observed, 
as required by supersymmetry. This depends on a set of Consistency Relations. The CR are
differences between regularization dependent integrals of the same degree of divergence 
that must be sett to zero. They are derived by the requirement that IR must be compatible 
with shifts in the loop momenta.

The CR were first obtained by constraining the Feynman amplitudes to be momentum routing invariant. 
As a consequence, this assures gauge invariance. An example was discussed in section II. Also in section 
II we have seen that setting the CR to zero is not the unique choice that implements gauge invariance. 
It is possible to get this by finding, for each specific amplitude, a relation between the local 
arbitrary constants that parametrize the Consistency Relations. Nevertheless, this relation would be 
different for different situations. This was the case we have faced when calculating the $(g-2)_l$ 
factor for SUGRA-QED: we could respect susy by choosing $\alpha_3=-35\alpha_2$, but this relation is not 
compatible with, for instance, the manifest gauge invariance of the vacuum polarization tensor for QED 
discussed in section II. Thus, the constrained version of Implicit Regularization ($\alpha_i=0$)
is the direct and natural way for obtaining supersymmetric and gauge invariant amplitudes.

Although we have verified it in a particular case, we believe that Implicit Regularization 
is supersymmetric invariant. The reason is that the rules of the constrained version of 
IR were tailored in such a way that symmetries that are manifest in the Lagrangian, and therefore 
in the untouched amplitude, are not spoiled. This is done by extending the properties of 
regular integrals to the regularized ones. These properties permits: shifts in the momentum of integration; 
cancellation between factors of the numerator and the denominator; algebraic manipulations in the integrand. 
This is in the same spirit of the Constrained Differential Regularization and Dimensional Regularization.
We can justify our procedure of giving up surface terms by arguing that it corresponds to automatically 
addind local symmetry restoring counterterms.

The results obtained in the present work were compared with previous calculations performed with Differential Regularization and Dimensional Reduction. It was found that the results are equivalent. In DRed the $\gamma$-matrix algebra is performed in four dimensions and the subsequent use of dimensional regularization automatically sets all regularization dependent (arbitrary) parameters to zero. This is explained by the fact that in 
Dimensional Regularization all the surface terms are null.
Therefore this method respects the CR and we can say that the two procedures will always yield identical results at one-loop order. Concerning the Constrained Differential Regularization, there are some evidences that the Consistency Relations are the momentum-space version of some of its rules. In a work in progress \cite{irdif}, the relation between the two techniques will be discussed.  

\section{Appendix}

We give bellow the on-shell results ($p^2=m^2$, $(p-p')^2=0$)
of the integrals which were necessary to the calculations 
of the previous sections:

\bq
I(p) &=& \int_k^\Lambda \frac{1}{[(p-k)^2-m^2]k^2} = I_{log}(\lambda^2) \nonumber \\ 
&+& b \ln{\left(\frac{\lambda^2}{m^2}\right)} +2b,
\eq
\bq
I_\mu(p)&=&\int_k^\Lambda \frac{k_\mu}{[(p-k)^2-m^2]k^2}= 
\frac{p_\mu}{2} \left \{  I_{log}(\lambda^2) \right. \nonumber \\
&+& \left. b \ln{\left(\frac{\lambda^2}{m^2}\right)}+b + \alpha_2\right\},
\eq
\bq
I_{\mu \nu}(p)&=& \int_k^\Lambda \frac{k_\mu k_\nu}{[(p-k)^2-m^2]k^2}= 
\frac 13 p_\mu p_\nu \left\{
I_{log}(\lambda^2) \right. \nonumber \\
&+& \left.b \ln{\left(\frac{\lambda^2}{m^2}\right)}
+ \frac 23 b  +\alpha_3\right\} + \mbox{$\eta_{\mu \nu}$ terms},
\eq
\bq
I(p,p') &=& \int_k^\Lambda \frac{1}{[(p-k)^2-m^2][(p'-k)^2-m^2]} \nonumber \\
&=& I_{log}(\lambda^2) 
+ b \ln{\left(\frac{\lambda^2}{m^2}\right)},
\eq
\bq
I_\mu(p,p')&=&\int_k^\Lambda \frac{k_\mu}{[(p-k)^2-m^2][(p'-k)^2-m^2]} \nonumber \\
&=& \frac{(p+p')_\mu}{2} \left \{  I_{log}(\lambda^2) 
+ b \ln{\left(\frac{\lambda^2}{m^2}\right)} \right. \nonumber \\
&+& \left. \alpha_2\right\},
\eq
\bq
I_{\mu \nu}(p,p')&=& \int_k^\Lambda \frac{k_\mu k_\nu}{[(p-k)^2-m^2][(p'-k)^2-m^2]} \nonumber \\
&=& \frac 16 (2 p_\mu p_\nu +2p'_\mu p'_\nu+p_\mu p'_\nu+ p'_\mu p_\nu) \left\{
I_{log}(\lambda^2) \right. \nonumber \\
&+& \left.b \ln{\left(\frac{\lambda^2}{m^2}\right)}
+\alpha_3\right\} + \mbox{$\eta_{\mu \nu}$ terms},
\eq
\bq
J&=& \int_k^\Lambda \frac{1}{[(p-k)^2-m^2][(p'-k)^2-m^2]k^2} \nonumber \\
&=&-\frac {b}{m^2} \int_0^1 dx \int_0^{1-x}dy \frac{1}{(x+y)^2},
\eq
\bq
J_\mu&=& \int_k^\Lambda \frac{k_\mu}{[(p-k)^2-m^2][(p'-k)^2-m^2]k^2} \nonumber \\
&=&-\frac{b}{m^2}(p+p')\mu,
\eq
\bq
J_{\mu \nu}&=& \int_k^\Lambda \frac{k_\mu k_\nu}{[(p-k)^2-m^2][(p'-k)^2-m^2]k^2} \nonumber \\
&=& -\frac{b}{6m^2}\left[ p_\mu p_\nu p'_\mu p'_\nu +\frac 12 (p_\mu p'_\nu +p'_\mu p_\nu) 
\right] \nonumber \\
&+& \mbox{$\eta_{\mu \nu}$ terms},
\eq
\bq
J'&=& \int_k^\Lambda \frac{1}{(p-k)^2(p'-k)^2[k^2-m^2]} \nonumber \\
&=&-\frac {b}{m^2} \int_0^1 dx \int_0^{1-x}dy \frac{1}{(x+y-1)^2},
\eq
\bq
J'_\mu&=& \int_k^\Lambda \frac{k_\mu}{(p-k)^2(p'-k)^2[k^2-m^2]} \nonumber \\
&=&-\frac{b}{m^2}\int_0^1 dx \int_0^{1-x}dy \frac{x(p+p')_\mu}{(x+y-1)^2},
\eq
\bq
J'_{\mu \nu}&=& \int_k^\Lambda \frac{k_\mu k_\nu}{(p-k)^2(p'-k)^2[k^2-m^2]} \nonumber \\
&=& -\frac{b}{m^2}\int_0^1 dx \int_0^{1-x}dy \frac{1}{(x+y-1)^2} \times
 \nonumber \\
&& \left[ (p_\mu p_\nu+p'_\mu p'_\nu)x^2+ \frac 12(p_\mu p'_\nu +p'_\mu p_\nu)xy\right]
\nonumber \\
&&+ \mbox{$\eta_{\mu \nu}$ terms}.
\eq
In the equations above, the on-shell integrals $J$, $J'$, $J'_\mu$ and $J'_{\mu \nu}$ are infrared divergent. 
They appear in the diagrams $\Lambda^{(4)}_\mu$, 
$\Lambda^{(5)}_\mu$, $\Lambda^{(9)}_\mu$ and $\Lambda^{(10)}_\mu$ and these singularities are cancelled pair by 
pair.

\section*{Acknowledgments}

MS and A. P. Ba\^eta Scarpelli thank CAPES-Brazil for financial support.


\begin{thebibliography}{88}

\bibitem{1} F. A. Berends and R. Gastmans, \textit{{Phys. Lett. \textbf{B}}} \textbf{55},  
311 (1975)

\bibitem{2} S. Ferrara and E. Remiddi, \textit{{Phys. Lett. \textbf{B}}} \textbf{53}, 
347 (1974)

\bibitem{3a} S. Ferrara and M. Porrati, \textit{{Phys. Lett. \textbf{B}}} \textbf{288},
85 (1992)

\bibitem{4a} S. Ferrara and A. Masiero,  Proc. of Susy 93, Northeastern University, 
Boston (1993); Bilchak , R. Gastmans and A. van Proeyen, 
\textit{{Nucl. Phys. \textbf{B}}} \textbf{273}, 46 (1986); A. Cullati, 
\textit{{Z. Phys. \textbf{c}}} \textbf{65}, 537 (1995)   

\bibitem{dimr} G. 't Hooft and M. Veltman, \textit{{Nucl. Phys. \textbf{B}}}, \textbf{44}, 
189 (1972)

\bibitem{3DR} F. del Aguila, A. M\'endez and F. X. Orteu, 
\textit{{Phys. Lett. \textbf{B}}} \textbf{145}, 70 (1984)

\bibitem{3} S. Bellucci, H. Cheng and S. Deser \textit{{Nucl. Phys. \textbf{B}}} 
\textbf{252}, 389 (1985)

\bibitem{3i} W. Wilcox, \textit{{Ann. Phys.}} \textbf{139}, 48 (1982)

\bibitem{3ii} M. T. Grisaru and D. Zanon, \textit{{Class. Quant. Grav.}} \textbf{2}, 477 (1985)

\bibitem{dimred} W. Siegel, \textit{{Phys. Lett. \textbf{B}}} \textbf{84}, 193 (1979); 
D. M. Capper, D. R. T. Jones and P. van Nieuwenhuizen, \textit{{Nucl. Phys. \textbf{B}}} 
\textbf{167}, 479 (1980)

\bibitem{4} Dominik St\"ockinger, hep-ph/0602005

\bibitem{1d} D. Z. Freedman, K. Johnson and J. I. Latorre, \textit{{Nucl. Phys. \textbf{B}}} 
\textbf{371}, 353 (1992)

\bibitem{2d} M. P\'erez Victoria, \textit{{Phys. Lett. \textbf{B}}} \textbf{442}, 315 (1998)

\bibitem{3d} F. del Aguila, A. Culatti, R. Mu\~noz Tapia and M. P\'erez Victoria, 
\textit{{Nucl. Phys. \textbf{B}}} 
\textbf{504}, 532 (1997)

\bibitem{4d} F. del Aguila, A. Culatti, R. Mu\~noz Tapia and M. P\'erez Victoria, 
hep-ph/9711474 

\bibitem{5d} F. del Aguila, A. Culatti, R. Mu\~noz Tapia and M. P\'erez Victoria, 
\textit{{Phys. Lett. \textbf{B}}} \textbf{419}, 263 (1998)

\bibitem{6d} F. del Aguila, A. Culatti, R. Mu\~noz Tapia and M. P\'erez Victoria,
\textit{{Nucl. Phys. \textbf{B}}} 
\textbf{537}, 561 (1999)

\bibitem{7d} F. del Aguila, M. P\'erez Victoria, hep-ph/9901291

\bibitem{6a} O. A. Battistel, {\it PhD thesis}, Federal University of 
Minas Gerais (2000)

\bibitem{6b} O. A. Battistel, A. L. Mota, M. C. Nemes \textit{{Mod. 
Phys. Lett. \textbf{A}}}
\textbf{13} 1597 (1998)

\bibitem{7}  A. P. Ba\^{e}ta Scarpelli, M. Sampaio and M. C. Nemes, 
\textit{%
{Phys. Rev. \textbf{D}}} \textbf{63}, 046004 (2001)

\bibitem{8}  A. P. Ba\^{e}ta Scarpelli, M. Sampaio, B. Hiller and M. 
C.
Nemes, \textit{{Phys. Rev. \textbf{D}}} \textbf{64}, 046013 (2001)

\bibitem {9} M. Sampaio, A. P. Ba\^eta Scarpelli, B. Hiller, A. 
Brizola,
M. C. Nemes and S. Gobira, \textit{%
{Phys. Rev. \textbf{D}}} \textbf{65}, 125023 (2002)

\bibitem{10} S. R. Gobira and M. C. Nemes, \textit{ Int. J. Theor. 
Phys.} \textbf{42},
2765 (2003)

\bibitem{11} D. Carneiro, A. P. Ba\^eta Scarpelli, M. Sampaio and M. C. 
Nemes,
\textit{JHEP} \textbf{12}, 044 (2003)

\bibitem{12} M. Sampaio, A. P. Ba\^eta Scarpelli, J. E. Ottoni, M. C. 
Nemes,
{\it Implicit regularization and Renormalization of QCD}, 
hep-th/0509102, to appear in the International Journal of Theoretical Physics

\bibitem{12i} Leonardo A.M. Souza, Marcos Sampaio, M.C. Nemes, 
\textit{{Phys. Lett. \textbf{B}}} \textbf{632}, 717 (2006) 

\bibitem{13} Carlos. R. Pontes, A. P. Ba\^eta Scarpelli, Marcos Sampaio, M. C. 
Nemes,
{\it Implicit regularization of massless theories}, 
hep-th/0605116

\bibitem{eds} E. W. Dias, A. P. Ba\^eta Scarpelli, Marcos Sampaio, J. L. Acebal 
and M. C. Nemes, work in progress

\bibitem{ABJ} S. Adler, \textit{{Phys. Rev.}} \textbf{177}, 2426 (1969); 
J. S. Bell and R. Jackiw, \textit{{Nuovo Cimento}} \textbf{51}, 47 (1969)

\bibitem{matgra} S. Ferrara, J. Scherk and P. van Nieuwenhuizen, 
\textit{{Phys. Rev. Lett.}} \textbf{37}, 1035 (1976); 
S. Ferrara et al., \textit{{Nucl. Phys. \textbf{B}}} 
\textbf{117}, 333 (1976); 
A. Dass, M. Fischler and M. Rocek, \textit{{Phys. Lett. \textbf{B}}} \textbf{69}, 1866 (1977); 
D. Z. Freedman and J. H. Schwars, \textit{{Phys. Rev. \textbf{D}}} \textbf{15}, 1007 (1977); 
E. Cremmer et al., \textit{{Nucl. Phys. \textbf{B}}} 
\textbf{147}, 105 (1979); 
E. Cremmer et al., \textit{{Phys. Lett. \textbf{B}}} 
\textbf{116}, 231 (1982)

\bibitem{irdif} Carlos R. Pontes, A. P. Ba\^eta Scarpelli, Marcos Sampaio, J. L. Acebal and 
M. C. Nemes, work in progress


\end{thebibliography}
\end{document}